# MATGANIP: Learning to Discover the Structure-Property Relationship in Perovskites with Generative Adversarial Networks

Junjie Hu[a,c,d], Mu Li[b], Peng Gao[a,c,*]

[a]*CAS Key Laboratory of Design and Assembly of Functional Nanostructures, Fujian Institute of Research on the Structure of Matter, Chinese Academy of Sciences, Fuzhou, Fujian 350002, P. R. China.*
[b]*Institute of Software, Chinese Academy of Sciences, Beijing, 100190, P. R. China*
[c]*Laboratory of Advanced Functional Materials, Xiamen Institute of Rare Earth Materials, Haixi Institute, Chinese Academy of Sciences, Xiamen 361021, P. R. China* [d]*University of Chinese Academy of Sciences, Beijing 100049, P. R. China*

**Abstract**

Accelerating the design of materials with artificial neural network draws more attention due to its magnitude potential. In the past works, some tools of materials information have been developed to promote the industrialize of state-of-the-art materials, such as the materials project, AFlow, and open quantum materials database. Else, more various endeavors are required for artificial general intelligence in the area of materials. In our works, we design neural networks named MATGANIP, which applies a combination of generative adversarial networks, graph networks, convolutional neural networks, and long short term memory for the perovskite materials. We adopt it for the building of a structure-property relationship, where the trained properties contain: the computational geometric property, tolerance factor; and the ground state property of Quantum theory, the vacuum energy. Moreover, the data-set about the $ABX_3$ perovskites is used for the learning of tolerance factor, to extend its function of tolerance factor into the structural identification of the arrangement of disorder atoms; another data-set about the density functional theory (DFT) calculation results is for the ground state energy of quantum theory, to obtain the more accurate result than the training of DFT calculated results in similar works. In our training task of the DFT date-set, we adopt the intuitive criterion to evaluate its performance: the mean absolute error (MAE) is smaller than 0.3 meV/atom, and the mean absolute error rate (MAER) is smaller than 0.01%. These results prove the potential ability of our MATGANIP in developing the relationship between materials structure and their properties. Notably, it suits for some scenes involved massive possible structures and their properties, such as the arrangement of the atoms in the structural phase transition of the inorganic perovskites with mixed atoms.

**Introduction**

Artificial neural networks (ANN) could play a significant role in the design of new materials.[1, 2, 3] Although the achievement of a critical material including the lab-to-fab transfer are usually time and cost-consuming, the high throughput DFT calculation can help to shorten the R&D period and accelerate the development of materials by quickly building up the quantitative materials-structure-property (QMSP) relationship.[4, 5, 6, 7] An immediate example is the fast developing quantum chemical approximations such as density functional theory (DFT)[8] facilitated mechanism rationalization of perovskite materials with matchless all-round optoelectronic properties.[9, 10, 11, 12, 13, 14, 15, 16, 17] Besides its specialty in predicting the pros and cons of a specific perovskite material and/or related devices, high-throughput computational screening become a hot topic for customized material design and screening.[18, 19] In this regards, the quantum chemical approximation find its opportunities in the search for lead-free and more stable perovskites[20, 21, 22, 23, 24] and other types of photovoltaic materials beyond the current $ABX_3$ type perovskites[25]. In the latter case, the double perovskites with the unit cell twice that of $ABX_3$ perovskite and two cations ordered on the B site[25] are currently under focus of many groups.[26, 27] The change of atomic composition will inevitably lead to the alteration of structure and phases, of which the stability cannot be entirely predicated by the traditional Goldschmidt's tolerance factor rule (equation 1), especially when facing the fact that the atoms are randomly distributed in their corresponding positions. [28, 29, 30, 31] To tackle the problem, it is necessary to establish a new structural factor based on the whole crystal structure instead of the mean radius of B-site atoms, which should also keep a correlation with Gibbs free energy.[32, 33]

To assist the experimental work of such star material, a series of high throughput screening, aiming at ascertaining the QMSP relationship while lessening the dependence on the trial-and-error experiments have been developed with accurate predictions.[20, 34, 35] However, due to the miscellaneous elemental doping engineering and fabrication approaches, the possibility of the perovskite derivative compositions and their subtle influence on the properties reaches an astronomical magnitude.[36, 37] The DFT based high throughput calculations have to be performed on enumerable samples which will require a vast amount of computing resources and make the search for new materials inefficient. It, therefore, becomes imperative to find a way to tackle the enormous amount of data out of the compositionally engineered perovskites. In this context, the immense potential of the machine learning models using artificial intelligence (AI) algorithms such as ANN could play a significant role with much-increased efficiency in processing the sophisticated data mining and material design of the off-stoichiometry perovskite materials.[37, 38, 39]

AI technology combines the powerful information processing capability of the computational server with the cognitively inferential capability of the enlightened human being.[40] Appropriate integration of different machine learning models based on a thorough understanding of their principles can promote the contribution of such promising methods to the development of material science.[41, 42, 43, 44, 45, 46, 47] To date, deep learning approaches like convolutional neural networks (CNN) and recurrent neural networks (RNN) are best used in the field of image processing, face recognition and voice recognition owing to the ability of effective transforming of edge details into global features.[48, 49] To solve the gradients vanishing problem of RNN, the long short-term memory (LSTM) module capable of learning long-term dependencies was also introduced.[50, 51] So far, attempts to extend the mathematical principle behind these algorithms out of their specified realms showed prospective potentials in improving the efficiency of analysis and prediction of QMSP relationships from the data collected in material science.[52, 2] Among models used to date, the generative adversarial nets (GAN)[53] featuring a system of two neural networks contesting with each other in a zero-sum game framework has been used in unsupervised machine learning and caught the attention of many researchers.[44, 53, 54] For example, Kadurin et al. implemented deep generative adversarial autoencoder to identify new molecular fingerprints with predefined anticancer properties.[55] Using the deep generative models, they found that the proposed generative adversarial autoencoder model significantly enhances the capacity and efficiency of development of the new molecules with specific anticancer properties.[56] Nouira et al. proposed a novel GAN called CrystalGAN which generates new chemically stable crystallographic structures with increased domain complexity.[57]

To simultaneously increase the materials learning efficiency, prediction accuracy and interpretability of an ANN in material science, in this work, we introduce an architecture called MATGANIP (Materials Generative Adversarial Network in Perovskites) based on the GAN model to establish QMSP relationship of complicated perovskite materials. Herein, the graph networks and CNN adopted in the MATGANIP provides an approach to extract the structural features for the generator; the LSTM contained in the MATGANIP improves its identification of the distribution for the discriminator. The inner layers in the generator and discriminator are used to enhance the representation of the structural features for the models interpretability and balance the ability between the generator and discriminator for increasing learning performance. To that end, we test the ability of MATGANIP in developing the QMSP relationships for two tasks: the structural tolerance factor of perovskites (denoted as MATGANIP.f) and the ground state energy of perovskites by DFT calculation (denoted as MATGANIP.e), respectively. Specifically, the structural tolerance factor of perovskites contains the geometric property of the perovskite crystal structure, which is widely used for the screening of stable cubic perovskite.[58] Therefore, the experiment with the learning of the tolerance factor (MATGANIP.f) can directly test the ability of MATGANIP in identifying the structural features. This learning results about the tolerance factor show that MATGANIP.f can build a similar mapping to the tolerance factor, from the generated features instead of atomic radii. Thus the application of MATGANIP.f depend on its input features, which contain more structural information of doped perovskites than atomic radii and the mean radii of mixed atoms.[23] The data-set used in the learning of the tolerance factor is extracted from the inorganic crystal structure database (ICSD).[59, 60]

In the meantime, we used ANN to learn the relationship between the structural features and the total energy calculated by DFT method. (denoted as MATGANIP.e) Actually, we could regard this experiment as a similar process to the calculation of quantum equation in the DFT software Firstly, we try to confirm that the learning result in MATGANIP.f contains valid structural information from the related perovskite, based on the test in MATGANIP.e. Then, in the DFT calculation step, the electronic wave function is used to describe the Hamiltonian of matter system relating to the space position of atoms and the numerical atomic potential of elements.[61, 62] Different from previous works about the machine learning of atomic potential,[63, 64] we put our focus on the machine learning of the space arrangement of atoms in the lattice. Since machine learning is always a data-driven process, we also think the current state of the interpretability of MATGANIP is not enough to replace the strict physical inference and mathematical approximation in the DFT calculations. In spite of this, the process of MATGANIP.e maybe provides access to understanding the learning of the ground state property behind the layers of ANN, instead of making the learning process become a black-box.

Moreover, the running efficiency of machine learning with high predictive accuracy has a significant advantage over the DFT calculation of large quantity of samples, which has been proved in past works.[65] In this case, the MATGANIP.e process may provide an access to the nature behind machine learning. The data-set used in MATGANIP.e is managed by the python with SQLite3.[66] The data-driven training in the MATGANIP.e could obtain a tiny predictive error with a small data-set. Hence, We prove that the MATGANIP strategy is as efficient as the strict quantum theory and has advantages over DFT calculation in simulating the property of inorganic crystalline materials with massive similar structures.

## Results and Discussion

### The Architecture of the MATGANIP

We build the MATGANIP, based on the framework of GAN, to find the relationship between the perovskites structure and its property. As shown in the diagram of MATGANIP model (figure 1), the structure data and property data are the two input of MATGANIP. The model weights were adjusted by iteration with backpropagation calculation via Adam optimizer[67], before it could reach a balance between the generator and the discriminator as the global optimum. The well-trained MATGANIP can provide the generator resulting from the learning model, which could predict the property data based on the input of structure data with small error. The inner layers of generator contain two CNN layers (Conv1 and Conv2) and a Linear layer. Every CNN layer is made up of a convolution layer (Conv2d), an activation function layer (ReLU)[68, 69], and a max-pooling layer[70] (MaxPool2d). Each of the three different layers is an indispensable part of CNN. They work together to grasp the features of input structure data. After the calculation by two CNN layers and a Linear layer, the structural data are transformed into the format of the property data suitable for the input of discriminator.

The organization of the discriminator sequentially includes the LSTM layer, Linear layer, ReLU layers, Linear layer, and sigmoid function layer.Under the framework of GAN, the discriminator is asked to identify the distributions of the data. During the training process of discriminator by iteration, the capability of it gradually increases, which can help the generator to build the relationship between the structure data and property data. The LSTM layer has a function to reinforce the ability of identification. The sigmoid layer[71], in the end, gives the output of discriminator in the format of probability value and provides an interface for the calculation of gradient. The other three layers work as transitive layers and adjust the performance of discriminator. At the last step of per iteration, backpropagation is performed according to the loss function and the result of which will be transferred to the generator (G) and discriminator (D), respectively. The whole work flow includes an iteration of the input of structural data with the backpropagation to D and G, and another iteration of the input of property data with the backpropagation to D. When the output of discriminator is equal to 0.5, the MATGANIP finishes the learning process.[53]

**Perovskites Represented by Graph Networks**

Graph convolutional network (GCN)[72] is a powerful neural network that leverages graph-structured data with convolutional networks. To make use of the feature of CNN, there was one study that used the PXRD graph patterns directly as the input data by treating them as bitmap and recognizing the pixels therein.[41] However, in this case, the internal periodic crystal structure information or knowledge, the information of the elements and their relationships cannot be reflected. In the meantime, ANNs like CrystalGAN[57] and CGCNN[42] for machine learning of material science, treated the crystal structure of a material as a finite graph composing nodes and edges that are denoting atoms and bonds or Euclidean distance respectively to describe the periodic lattice. The pitfalls of this approach are that the graph-structure data cannot be correctly linked to the periodicity of the crystal structure, and error between the models and the real materials may generate at the boundary of lattice and graph. For example, the stoichiometric ratio represented in the graph-structure data is not equal to the actual formula of inorganic materials. It has been concluded that the graph composed directly of atoms and bonds is more suitable for organic molecule, rather than inorganic materials.[73]

To circumvent the problems related to the periodicity and boundary of inorganic crystal lattice, in our study, we use the graph-structured data for the representation of the original PXRD data, which is in the Brillouin zone[74] of the reciprocal space. Because the reciprocal space is transformed via Fourier transformation from the real space of the periodic crystal structure, the periodic crystal plane is equal to a point in the reciprocal space that is corresponding to the peak in the PXRD pattern. Specifically, each of the diffraction peaks in PXRD data corresponds to a data-pair formed by a diffraction angle ($\theta^* = 2 * \theta$) and an intensity ($I$) value. Together they contain the structure information including atomic coordinates, the correlation of different atoms, and the information of the elements. As shown in figure 2 left, significant diffractions containing the data-pair in the PXRD pattern of perovskite structure were chosen as the feature data. If we consider each peak as a vertex and the relationship between each two linked peaks as an edge, all the vertices and edges can make up of the graph network used in the MATGANIP, which could be used to represent the original PXRD pattern. We show the typical visualized graph network in the middle of figure 2. All of the vertices and edges are organized in the format of graph networks and marked with gradient colors. Each point corresponds to a diffraction peak. By selecting the peaks as the nodes of the graph networks and the difference of peaks as the edge, we could avoid the error reduced by the boundary. After that, these graph-structured data can be received by the CNN layers in MATGANIP following the logic of GCN. (figure 2 right) The process of transforming the XRD pattern of perovskites into the format of graph-structure data is one of the most critical and innovative steps in the execution of MATGANIP. Compared with previous methods, our strategy can build a direct correlation between the features needed in MATGANIP and the most important structural information in perovskite crystal.

**The Data-flow of the MATGANIP**

A reliable neural networks model brings together intelligent algorithm design and data. In the above sections, we have discussed the ideas of the architecture and input data preparation of the MATGANIP. The application of MATGANIP also depends on the data-set. In order to verify its ability to develop the relationship between the structure and the property, two different data-sets and related QMSP relationships were used: 1. Crystal structure data of $ABX_3$ perovskites from ICSD and their tolerance factors as the property data (MATGANIP.f); 2. The derivative structures of $CsPbI_{3-x}Br_x$ ($x = 0.375$) and these structures Gibbs free energy at the vacuum state by DFT calculation (MATGANIP.e), as shown in figure 3 with blue and grey respectively.

In figure 3, the data-flow of MATGANIP.f is signed by the color of light blue. Similarly, the gray arrows indicate the data-flow of MATGANIP.e going in and out of SQLite3 database. The QMSP for the above MATGANIP.f refers to the crystal geometry structure represented by tolerance factor and our particular structure representation with the graph-structured data representing PXRD data. For MATGANIP.e, the QMSP relates to the ground state energy directly related to the structure and the crystal graph-structured representation. The process of graph-network by GCN (green part in figure 3) and reading of property data by LSTM layers (purple part in figure 3), are shared by different tasks in MATGANIP. The data processing of GCN provides the graph-structured representation for the crystal structure according to the Laplacian matrix, which elements are the inner product of the two-tuple vector of different diffraction peaks. For arbitrary $(\theta_i^*, I_i)$ and $(\theta_j^*, I_j)$, the matrix elements ($a_{ij}$) is equal to $\theta_i^* * \theta_j^* + I_i * I_j$. The above is the method of MATGANIP to digitalize the crystal features, which can provide unified algorithm for the learning of crystal materials.

As per the concept of GAN model, the logical process in LSTM layers is to identify the fake property data and the real property data. The real property data is from our materials knowledge or the calculated results based on the Quantum theory; on the other hand, the fake property data comes from the structural data handled by the generator. The discriminator continuously distinguishes the fake data from the real data until the point that the generator can provide fake data that is almost equal to the real data, then the MATGANIP could find the relationship between the structure data and property. The LSTM layers have a function to balance the ability of G and D by enhancing the identification of D.[75] Inside LSTM, the sub-layers calculates the internal parameters ($h$ and $c$) with the input property data signed by variable $x$. The LSTM results signed with variable y are transferred into the next layer of the discriminator. As it has been proved, the parameter $c$ is used to contain the long-scale features, and the parameter $h$ is used to keep the short-scale feature; and the ability to identify the distribution depends on the long-scale and short-scale features, which can improve the identification of D. Furthermore, the MATGANIP can optimize itself according to the backpropagation. If the training step is an odd number, the MATGANIP adjust the performance of discriminator for better recognition; otherwise, if the training step is even number, the MATGANIP can regulate the performance of generator to obtain more precise QMSP relationship while improving the discriminator.

**The Performance of the MATGANIP**

With the results about tolerance factor(MATGANIP.f) and DFT-ground-state energy (MATGANIP.e), the ability of MATGANIP has been verified to establish the QMsP relationship for the inorganic perovskites. The perovskites crystal materials have the prototype formula $ABX_3$, where the B-site atoms and X-site atoms coordinate to format the 3-dimensional $BX_6$ octahedron networks with sharing points connected, and the A-site atom exists the space between these octahedrons.[10] Under ideal conditions, each atom is equivalent to a rigid sphere, and the tolerance factor applies visible geometric characteristics for describing the structural features of perovskites. Therefore, during the learning of the perovskite crystal structure by MATGANIP.f, the knowledge about tolerance factor can directly verify the ability of MATGANIP in identifying the perovskites structures.

In figure 4, we give the training process of MATGANIP.f while evaluating its performance on both the test-set and train-set. According to the algorithm of GAN, the generator (G) of MATGANIP.f could synthesize the structural factor as the "fake" property data. Meanwhile, the discriminator (D) of MATGANIP.f could distinguish the distributions of the real property data ($x_r$) and fake" property data ($x_f$). The output of D ($D(x)$) is probability value from the Sigmoid layer of D. On the other hand, the loss functions of G and D are used to optimize their behaves. The feedback signals for optimization come from the gradient descent by Adam, where the MATGANIP.f calculates the loss function of D with $D(x_r)$ and $D(x_f)$ and the loss function of G with $D(x_f)$. (See Methods Section) During the iteration of the training step, MATGANIP.f gradually improve the identification of D and optimize the synthetic ability of G under the backpropagation. The MATGANIP reaches the global optimum just as both of $D(x_r)$ and $D(x_f)$ are equal to 0.5. Hence, the probability value of $D(x_f)$ and the loss function of D indicate enough reliability to describe the training of MATGANIP.f in figure 4a.

At the beginning of the training step in figure 4a, the tremor in the curve of D's loss function indicates that the correct identification of the input property data has not been obtained. As the curve of D's loss function turns flat, D generally gained the ability to distinguish the fake data and real data. According to the equation 2 in Methods section, we can deduce that the global optimum of loss function of D approximates to 1.3863 (-2$log$0.5). At the end of the training step, the curve of the loss function of D approached to this global optimum. Similarly, the curve of the value of $D(x_f)$ synchronously appears to oscillate at the initial stage of training. The fake property data comes from the G and the oscillation of $D(x_f)$ indicates that the training of G relies on the performance of D. Namely, when the value of $D(x_f)$ is beyond the global optimum line, it means that the fake data from G can successfully deceive the identification of D and MATGANIP.f needs to improve the ability of D. And when the value of $D(x_f)$ is below the global optimum line, an opposite situation is appeared. At the end of training, the value of $D(x_f)$ approaches to nearly 0.5, when the curve of D's loss function reaches the global optimum line, suggesting that the G has been well trained well along with the training of D and the MATGANIP.f finishes the training process. After the training of MATGANIP.f with the structural data of the prototype formula $ABX_3$ extracted from ICSD, we evaluate the MATGANIP with the mean absolute error (MAE) (figure 4b) and mean absolute error rate (MAER) (figure 4c) on the train-set and test-set. According to the results, we find that most of predictive results have absolute errors smaller than 0.05 and absolute error rates close to 5%. Meanwhile, the MAE and MAER are showing higher values close to 0.08 and

9 %, respectively. This is because the structure factor of MATGANIP.f is not limited by the tolerance factor defined by equation 1 anymore, and MATGANIP.f has extended the boundary of the concept. In other words, a similar but not the same definition of structure factor is created, which is different from previous works.[31, 30] The performance of MATGANIP.f will be discussed in the following sections.

The classic Goldschmidts tolerance factor relies on the radius of atoms from different sites. For the situation that one site may be occupied by either of two or more atoms, will be calculated based on the mean radii of these atoms.[23] MATGANIP.f is used to build a new relationship that can be used in the more complex and broad scenes between the PXRD data of perovskites and the tolerance factor. For example, in double perovskite there are a series of possible arrangements because two different atoms exists on the B-site in double perovskite. Furthermore, the tolerance factor based on the mean radius could only give the same value for different structures, but the factor by MATGANIP.f could differentiate the different structures. As is known, both of the Gibbs free energy and tolerance factor have an evident correlation with the structural stability of perovskites.[23] Since the tolerance factor cannot reflect the difference of the arrangement of mixed B-site atoms in double perovskites, we compare the structure factor by MATGANIP.f with the Gibbs free energy at the vacuum state to check the performance of MATGANIP.f. As shown in support information Table S1, we list the arrangements of atoms (B-site) disorder in the 2 * 2 * 2 super-lattice of double perovskite and their classification of point group symmetry According to the symmetry of B-site atoms, these possible derivative structures, relating to the arrangement of mixed B-site atoms, can be signed with Td, C3v, C4v, D2h, Cs and C2.[76] Herein, we use 15 different formula of Lead-free perovskites to analyze this structure factor of MATGANIP.f. These perovskites contain $Cs_2B^1B^2I_6$ where $B^1$ is Sn, and $B^2$ is Ca, Ti, V, Mn, Fe, Ni, Cu or Zn; also contain $Cs_2B^1B^2Br_6$ where $B^1$ is Ag, In or Tl, and $B^2$ is In, Tl, Sb, or Bi. Therefore, there are a total number of 90 perovskite structures in the heat-map of figure 5, inside which the results of MATGANIP.f are also compared with the Gibbs free energy of these structures. As shown in figure 5, each column corresponds to a type of perovskite formula; every row corresponds to a specific arrangement of disordered B-site atoms. In each colorful square, the up-triangle denotes the Gibbs free energy of each structure, and the down-triangle expresses its corresponding structure factor by MATGANIP.f. Their colors stand for the value, and the color bars are given in the right part of the heatmap. The deep red color represents the maximum of the structure factor by MATGANIP.f, which appears in the $Cs_2SnNiI_6$ with the arrangement of Td point group[76]. The deep blue color means the minimum of the structural factor by MATGANIP.f, which corresponds to the $Cs_2SnCaI_6$ with Td group. Meanwhile, $Cs_2SnCaI_6$ with D2h group, $Cs_2SnTlI_6$ with D2h group, $Cs_2SnNiI_6$ with Cs group, and $Cs_2SnCuI_6$ with Cs group are close to the minimum. On the other hand, the bright yellow color stands for the maximum of Gibbs free energy, which relates to the column of $Cs_2SnZnI_6$. The bright pink color indicates the minimum of Gibbs free energy, which concerns with the column of $Ca_2SnTiI_6$ and $Ca_2SnVI_6$. The upper-left color shows a broader variety between different columns than among the rows. Meanwhile, the lower-right color demonstrates a gradient tendency on the columns of each row and has a small difference among the rows. Generally, the structural factor by MATGANIP.f can give out the structure features, when the arrangement of disordered atoms is considered. Although the Gibbs free energy is also influenced by the atomic potential of different elements, we can confirm that the structural factor by

MATGANIP.f correlates with the Gibbs free energy for each perovskites, according to the correlation given by of the Pearson correlation coefficient in the supplement information figure S2.

When we consider the derivative structures relating to the arrangement of disorder atoms in the crystal lattice, the quantity of the possible structures of doped perovskites is usually more than it of the double perovskites and maybe reach an astronomical magnitude.[77, 36] Although it is significant to distinguish the difference of these structures and explore the relationship of their properties for accelerating the development of this type materials, it will be difficult due to the huge computing resource consumption. We noticed the potential of MATGANIP to achieve this mission, according to its performance in the learning of tolerance factor. Herein, we consider a data-set made up of 2024 possible structures, where the formula is $CsPbI_3$ doped by 12.5% Br with 2*2*2 super-lattice. The calculations of DFT-ground-state energy are running on these structures by VASP, and their results are stored in the SQLite3 database. Furthermore, MATGANIP.e applies this data-set for the training of DFT-ground-state energy, which is related to the ground state property of Quantum theory. As we have done in the MATGANIP.f, we also utilize the curve of $D(x_f)$ and the loss function of D to evaluate the training process and use the MAE and MAER to determine the final performance of MATGANIP.e on test-set and train-set respectively. As shown in figure 6a, the curve of loss function of D gradually converges to its optimal value of 1.3863 ($-2log0.5$), and the curve of $D(x_f)$ also reaches its optimal value with an error smaller than 0.0001. Both of the curves of $D(x_f)$ and loss function of D show a shake in the middle of training, which originates from a perturbation from learning rate. At the end of training, we finish the learning process of MATGANIP.e on the data-set of DFT-ground-state energy under the logic of GAN. According to the statistical data in figure 6b and 6c, the MAE of MATGANIP.e at the data-set of DFT-ground-state energy is smaller than 0.3 meV per atom, and the MAER of it is smaller than 0.01%. The precision of prediction confirms the MATGANIP.e has a big advantage on the discovery of QMSP relationship, especially among the doped materials with many derivative structures. In previous works[38], an much larger MAE of 20-34 meV per atom was achieved from an ANN model for the DFT calculated energy of perovskites. Although the data-set used in MATGANIP.e is different from their work, the effectiveness and precision of MATGANIP.e based on common criterion is proved.

Although the learning ability of MATGANIP.f and MATGANIP.e, we need to point out that the training of MATGANIP relies on some skills related to the understanding of the logic of GAN and gradient descent which was improved in the work about Wasserstein GAN.[78] Moreover, the random strategy of training samples is also limited the GAN, which could be enhanced by reinforcement learning. Because of these drawbacks, even if the running of MATGANIP nearly reached its optimum, the MAE of MATGANIP.e showed the distribution of error: the absolute error of some samples are more than for times of MAE, and too much samples are close to the MAE. We hope to make the distribution of predicted results sufficiently close to the real data, after adding to the frameworks of Wasserstein GAN and reinforcement learning.

In our works, we highlight the use of the graph-structured networks of the PXRD peaks to identify the structural features of perovskites, which is sensitive for the arrangement of disordered atoms in crystal lattice. Also, the MATGANIP provides a framework to build the QMSP relationship for perovskites. In the experiment with the data-set about the property of perovskites, we find that the MATGANIP for the tolerance factor (MATGANIP.f) can mimic and extend the Goldschmidts tolerance factor to get a new structure factor that can reflect the differences induced by the different arrangements of mixed atoms in the crystal lattice and have a direct correlation with the Gibbs free energy at the vacuum state. Hence, we confirmed the effectiveness of MATGANIP.f. Furthermore, we apply the MATGANIP for learning DFT-ground-state energy (MATGANIP.e), to develop the relationship between the structures of 2024 inorganic $ABX_3$ perovskites and the corresponding ground-state energy based on the DFT calculation. The results showed that the MATGANIP.e could predict the energy with a tiny error after the data-driven learning. This indicates that the function of MATGANIP.e is similar with DFT method to generate energy values from the crystal information, which could be regarded as another mode of machine learning for the quantum equation in the previous DeepMind work of organic molecule. Their works are made under the code framework of tensorflow,[79] and our code is based on the module of PyTorch.[80] Meanwhile, the architectures of machine learning models are different, albeit that our work lacks more strict inference with quantum equation compared with that in DeepMind.[81]

## Conclusion

In summary, we accomplished the ANN called MATGANIP, for the first time to effectively develop the QMSR relationship of inorganic perovskites materials based on the GAN and graph-structured PXRD data. MATGANIP.f learned the tolerance factors from the classic perovskite formula in ICSD with the prototype structure of $ABX_3$, and extend the Goldschmidts tolerance factor into a new application for the derivative structures related to the double perovskites and doped perovskites. Then, MATGANIP.e was trained on the data-set of the DFT-ground-state energy of doped perovskites and gave an accurate prediction of the DFT-ground-state energy for these structures with a low MAER of 0.01%. These shreds of evidences point out the MATGANIP is valid for the building of QMSP relationship of perovskites materials. Moreover, the logic of GCN contained in the MATGANIP regarding the perovskite structures shown as graph-structured data made the analyses the graph-structured networks by the layers of CNN possible, which enhanced the interpretability of the MATGANIP. Although the MATGANIP was only used to learn the perovskite properties relating to the empirical knowledge and the ground-state energy of Quantum theory, we believe that the MATGANIP can also support other properties at ground state, such as elastic constants, static dielectric and so on. GAN played a significant role in the MATGANIP by providing a way to discover the QMSP relationship. Next step, we should adopt other techniques to improve the current GAN model, such as Wasserstein GAN and reinforcement learning. Beyond its maiden attempt at learning inorganic perovskites materials, we believe the MATGANIP can endeavor to do more works for developing new inorganic-organic hybrid perovskites and low-dimensional perovskites.

## Methods

### Crystal structure information

The perovskite structures used in our works come from these with the prototype of $ABX_3$ in ICSD and the SQLite3 generated by traversing the derivative structures of doped perovskites.

**Tolerance factors**

The Goldschmidts tolerance factor of $ABX_3$ perovskite is calculated as the following formula:

$$t=\frac{r_A+r_x}{\sqrt{2}(r_B+r_x)} \qquad (1)$$

In equation 1, $r_A$, $r_B$, and $r_C$ are the ionic radii of A, B, X in the chemical formula. The process of calculation is running by the python with the Pymatgen package.

**The graph-structured matrix in GCN**

The results of PXRD are simulated by the Pymatgen package with CuKa radiation. The PXRD contains the information of different peaks. Each peak stands for a crystal face and has a format of a diffraction angle ($\theta^* = 2 *\theta$) and intensity ($I$). We use the inner product of peaks as the element of this matrix which corresponds to the Laplacian matrix. This matrix also stands for the graph networks which is a set of vertices and edges.

**The running of the MATGANIP**

In this work, all of the neural networks perform by the pytorch[80] with Adam optimizer at a learning rate of 0.01. The data-set used in the work was randomly split into train-set and test-set with a proportion of 80:20.

The MATGANIP searches the global optimum according to the primary algorithm of GAN. Moreover, for each perovskites, there are the $x_r$ extracted from the database, and $x_f$ generated by the G of MATGANIP. We update the MATGANIP with the loss function, which can be written as:

$$F_{loss}(D) =- ( \log D(x_r) + \log (1 - D(x_f) ) \qquad (2)$$
$$F_{loss}(G) = \log (1 - D(x_f)) \qquad (3)$$

The equation 2 is the loss function of D, and the equation 3 is the loss function of G. When both of $D(x_r)$ and $D(x_f)$ are equal to 0.5, the MATGANIP reaches the global optimum.

**DFT calculations**

All of the DFT calculations in this work were performed using the Vienna ab-initio simulation package (VASP) within the projector augmented-wave approach. The Perdew-Burke-Ernzehof generalized gradient approximation exchange-correlation functional and a plane-wave energy cut-off of 550 eV were used. The energy was converged to within $5 * 10^{-5}$ eV per atom and force within 0.01 eV/angstrom.

## The auto-work flow for the MATGANIP

We tackle the batch process of the running of VASP with the bash shell and python script and manage their results by the python with SQLite3.

## Acknowledgments

P.G. acknowledges the financial support from the National Natural Science Foundation of China (Grant No. 21975260) and the Recruitment Program of Global Experts (1000 Talents Plan) of China (Nos.201613, Y8d3071caa).

## Author contributions statement

P.G. conceived the work. J.H., M.L., and P.G. discussed and designed the experiments. J.H. and M.L. collected the data with the python and SQLite3. J.H. finished the code related to the MATGANIP, and the works about the DFT calculations and managed the running result with shell and python. J.H. and P.G. wrote the manuscript. J.H., M.L., and P.G. edited the manuscript. All authors reviewed the manuscript.

## Additional information

The authors declare no competing financial interests.

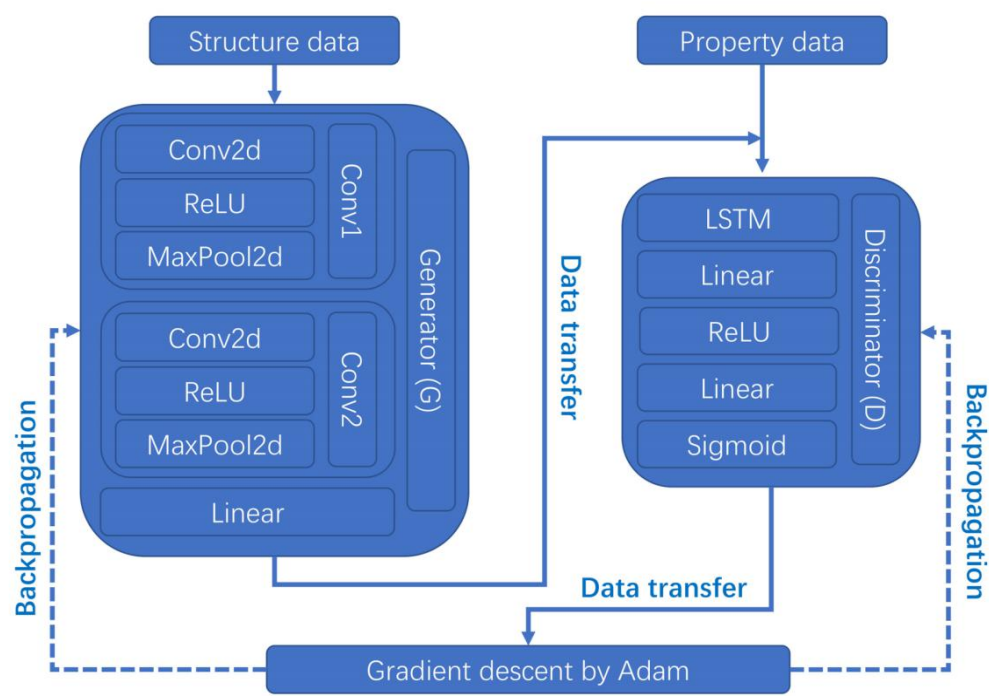

**Figure 1: The schema diagram of MATGANIP.** The structure data and property data are the input of the generator (G) and the discriminator (D) of the MATGANIP, respectively. In the inner of the G, there are the Conv1, Conv2, and Linear. Both of the Conv1 and Conv2 are the CNN layers, which contains the Conv2d, ReLU, and MaxPool2d layers. Linear is fully connected layers and provides a data-transfer from the G to the LSTM of the D. In the inner of the D. There are LSTM, Linear, ReLU, Linear, and Sigmoid. LSTM represents the long short term memory and enhances the ability to identify of the distribution of the input data. The sigmoid layer provides a Sigmoid function for the D and makes the output of the D to be a probability value. The other three layers, two Linear and ReLU, transform the data of LSTM into the Sigmoid. ReLU provides an active function. Based on the Adam, gradient descent calculation receives the data of the Sigmoid of the D and generates the backpropagation for the G and D of the MATGANIP.

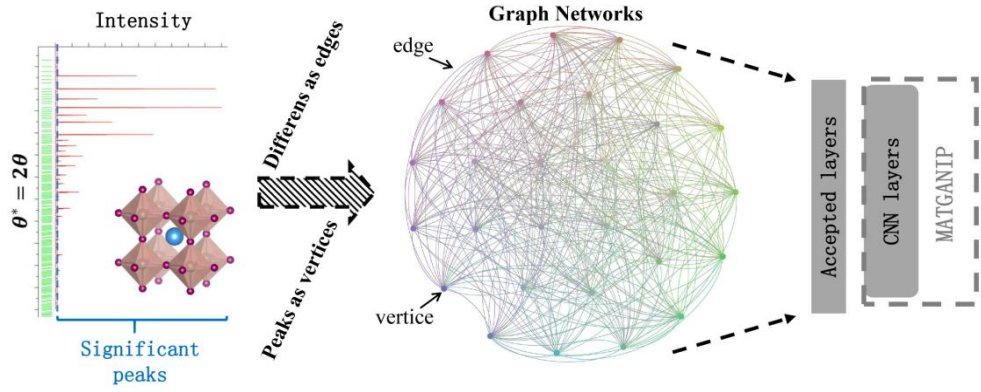

**Figure 2: The principle about that The GCN extracts the features of perovskites crystal structures.** At the left shows the process to transform the lattice of crystal perovskites into the PXRD data. The graph-structured data contains the PXRD peaks as vertices and differences of peaks as edges. In the middle shows the graph network made up of 28 different peaks, where the color lines represent the difference between two peaks. Gradient colors are used to sign the different vertices and the edges for ease of distinguishing. We transfer the graph networks into the CNN layers of the MATGANIP, which is listed on the right of this map.

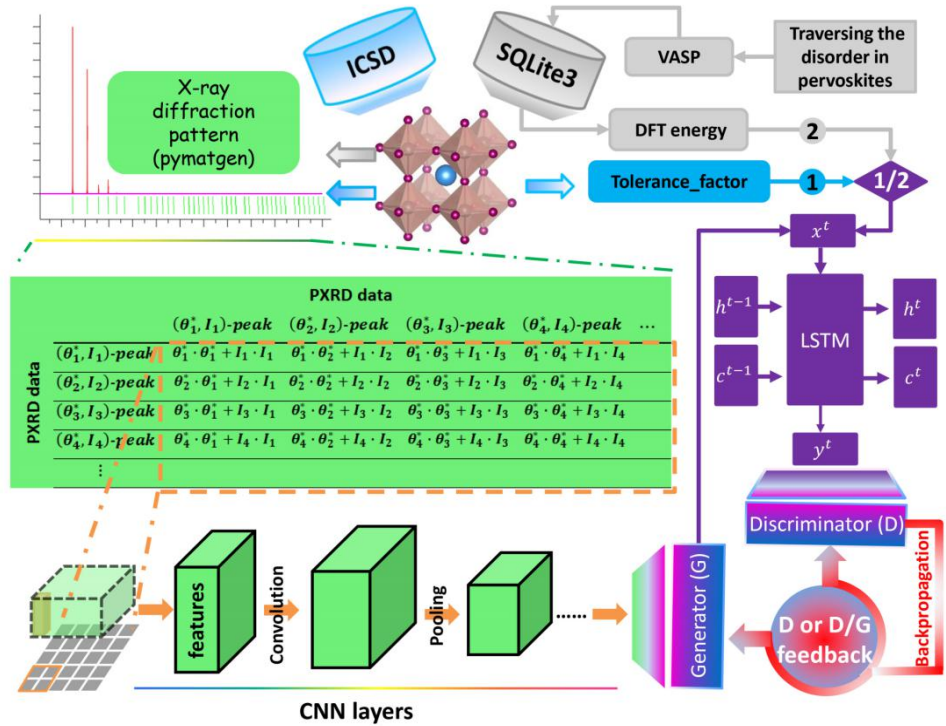

**Figure 3: The data flow of MATGANIP.** The light blue part in the map represents the management of data-set in the learning of MATGANIP.f; The grey part shows it in learning MATGANIP.e, respectively. The green part gives out the logic of GCN. At the top of green part shows the X-ray diffraction pattern data calculated by python with Pymatgen package; in the middle of the green part shows the inner product of peak data as the matrix elements of the matrix features; at the bottom of green part

shows the running of the CNN layers. The purple part indicates the running of long short term memory (LSTM) layers, where x stands for the input element, y represents the output elements of LSTM, and h and c denote the parameters of the inner of LSTM, respectively. At the bottom right corner, we point out the backpropagation of the MATGANIP, where the D generates value of the backpropagation and MATGANIP transfers the feedback signal to the D or both of the D and G.

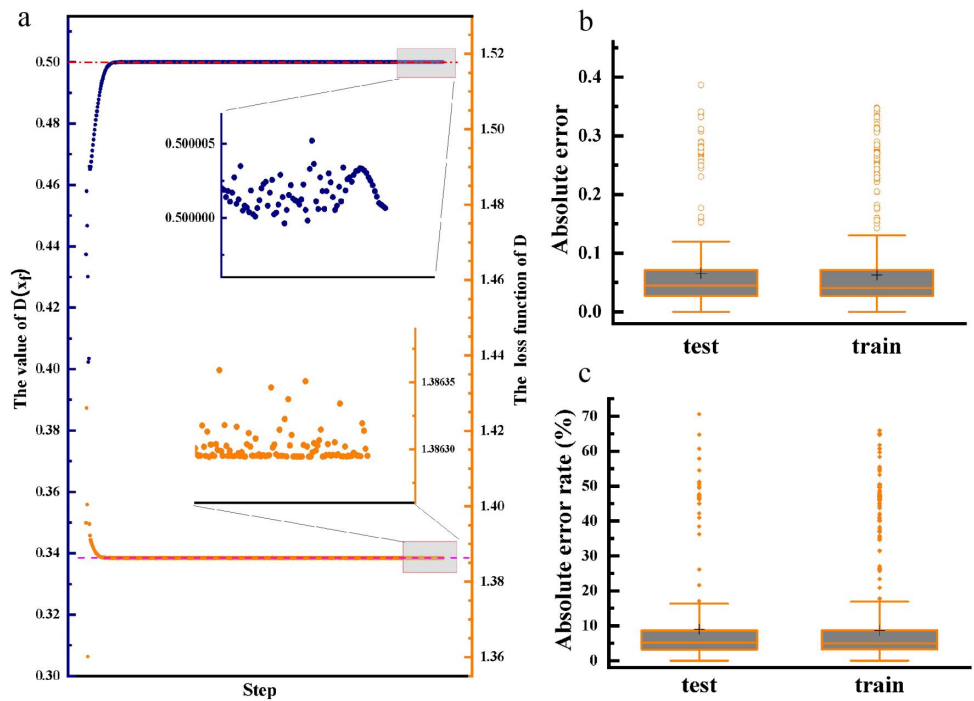

**Figure 4: The performance of the MATGANIP.f.** a). The curve of the value of $D(x_j)$ records the training process, with a global optimum value of 0.5; the curve of the loss function also records the training process, with a global optimum value of -2log0.5; the enlarged map of the end of both curves are given. b). The boxplot of the mean absolute error on the test set and train set, respectively. c). The boxplot of the mean absolute error rate in the test set and train set, respectively. In the b and c, the grey box stands for the interquartile range(IQR), and the range with 1.5 IQR is also given. The line in the box represents the median line. The cross model stands for the mean value. The open circle represents the outliers.

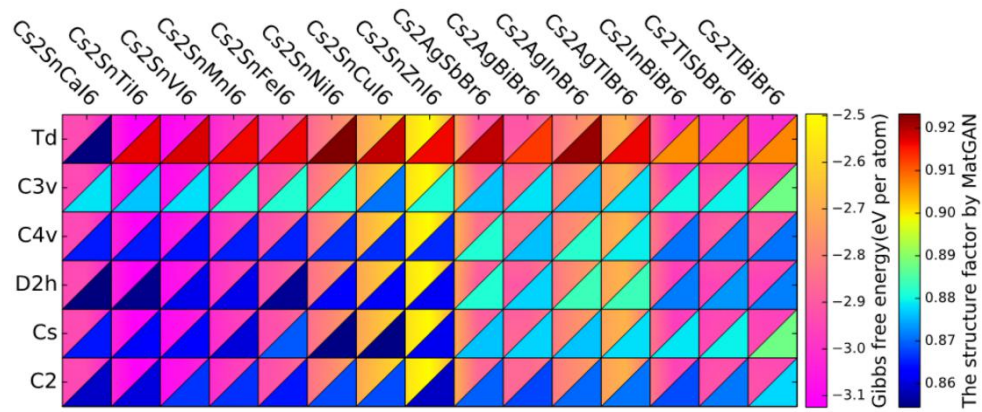

**Figure 5: The heatmap of the lead-free double perovskites.** At the top, we list the formula of these double perovskites. Shown are two of the descriptor parameters and their correlation, the Gibbs free energy and the calculated structure factors of MATGANIP.f, ordered with the point group of B-site mixed atoms in vertical direction and the atomic number of B-site mixed atoms in horizontal direction. Each square represents a structure; in each square, the upper-left triangle indicates the Gibbs free energy, with pink for smaller value and yellow for higher value, whereas the lower-right triangle indicates the structure factor of MATGANIP.f, with red for the value of 0.92 and the blue for the value of 0.86. The Pearson correlation coefficient of each row is given in support information.

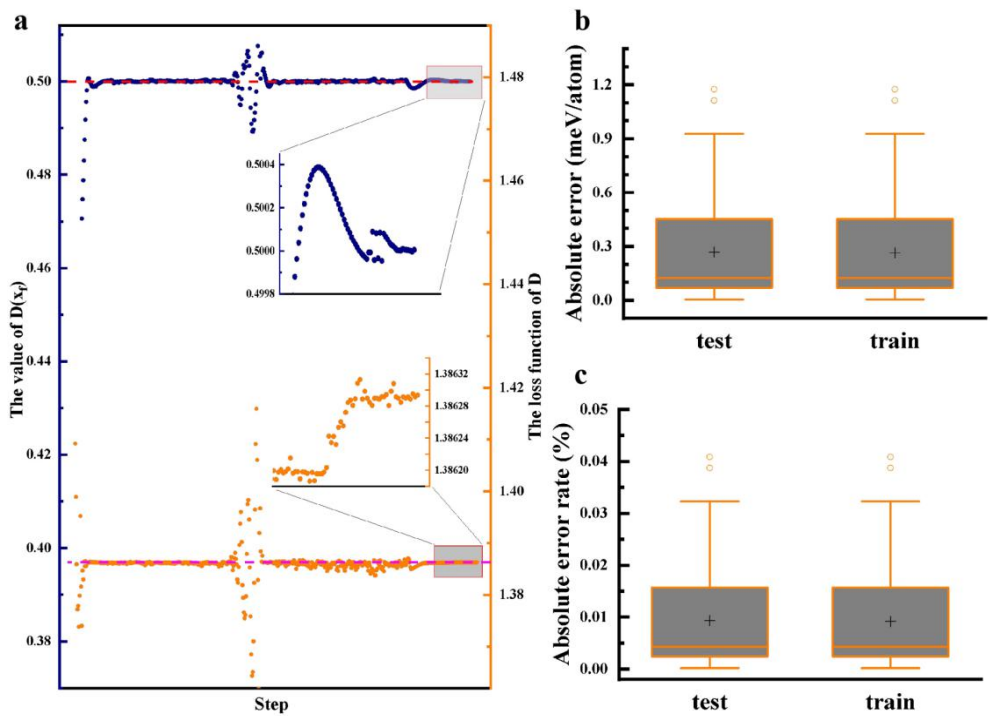

**Figure 6: The performance of the MATGANIP.e.** a). The curve of the value of D(xf) records the training process, with a global optimum value of 0.5; the curve of the loss function also records the training process, with a global optimum value of -2log0.5; the enlarged map of the end of both curves are given. b). The boxplot of the mean absolute error on the test set and train set, respectively. c). The boxplot of the mean absolute error rate in the test set and train set, respectively. In the b and c, the grey box stands for the interquartile range(IQR), and the range with 1.5 IQR is also given. The line in the box represents the median line. The cross model stands for the mean value. The open circle represents the outliers.

[*]*Corresponding authors at: Fujian Institute of Research on the Structure of Matter, Chinese Academy of Sciences, Fuzhou, Fujian 350002, P. R. China.*
*E-mail addresses: Peng Gao. (peng.gao@fjirsm.ac.cn).*